\documentclass[aps,twocolumn,prd,showpacs,showkeys,preprintnumbers,superscriptaddress,nobibnotes,floatfix,longbibliography,notitlepage,nofootinbib]{revtex4-2}

\usepackage{amsmath}
\usepackage{amsfonts}
\usepackage{amssymb}
\usepackage{graphicx}
\usepackage[colorlinks=true,allcolors=blue]{hyperref}
\usepackage[capitalize]{cleveref}
\usepackage[dvipsnames,table]{xcolor}
\usepackage{appendix}
\usepackage{lineno}
\usepackage{float}
\usepackage{afterpage}
\usepackage{placeins}

\begin{document}

\title{Limiting the Parameter Space for
Unstable eV-scale Neutrinos Using IceCube Data}

\affiliation{III. Physikalisches Institut, RWTH Aachen University, D-52056 Aachen, Germany}
\affiliation{Department of Physics, University of Adelaide, Adelaide, 5005, Australia}
\affiliation{Dept. of Physics and Astronomy, University of Alaska Anchorage, 3211 Providence Dr., Anchorage, AK 99508, USA}
\affiliation{School of Physics and Center for Relativistic Astrophysics, Georgia Institute of Technology, Atlanta, GA 30332, USA}
\affiliation{Dept. of Physics, Southern University, Baton Rouge, LA 70813, USA}
\affiliation{Dept. of Physics, University of California, Berkeley, CA 94720, USA}
\affiliation{Lawrence Berkeley National Laboratory, Berkeley, CA 94720, USA}
\affiliation{Institut f{\"u}r Physik, Humboldt-Universit{\"a}t zu Berlin, D-12489 Berlin, Germany}
\affiliation{Fakult{\"a}t f{\"u}r Physik {\&} Astronomie, Ruhr-Universit{\"a}t Bochum, D-44780 Bochum, Germany}
\affiliation{Universit{\'e} Libre de Bruxelles, Science Faculty CP230, B-1050 Brussels, Belgium}
\affiliation{Vrije Universiteit Brussel (VUB), Dienst ELEM, B-1050 Brussels, Belgium}
\affiliation{Dept. of Physics, Simon Fraser University, Burnaby, BC V5A 1S6, Canada}
\affiliation{Department of Physics and Laboratory for Particle Physics and Cosmology, Harvard University, Cambridge, MA 02138, USA}
\affiliation{Dept. of Physics, Massachusetts Institute of Technology, Cambridge, MA 02139, USA}
\affiliation{Dept. of Physics and The International Center for Hadron Astrophysics, Chiba University, Chiba 263-8522, Japan}
\affiliation{Department of Physics, Loyola University Chicago, Chicago, IL 60660, USA}
\affiliation{Dept. of Physics and Astronomy, University of Canterbury, Private Bag 4800, Christchurch, New Zealand}
\affiliation{Dept. of Physics, University of Maryland, College Park, MD 20742, USA}
\affiliation{Dept. of Astronomy, Ohio State University, Columbus, OH 43210, USA}
\affiliation{Dept. of Physics and Center for Cosmology and Astro-Particle Physics, Ohio State University, Columbus, OH 43210, USA}
\affiliation{Niels Bohr Institute, University of Copenhagen, DK-2100 Copenhagen, Denmark}
\affiliation{Dept. of Physics, TU Dortmund University, D-44221 Dortmund, Germany}
\affiliation{Dept. of Physics and Astronomy, Michigan State University, East Lansing, MI 48824, USA}
\affiliation{Dept. of Physics, University of Alberta, Edmonton, Alberta, T6G 2E1, Canada}
\affiliation{Erlangen Centre for Astroparticle Physics, Friedrich-Alexander-Universit{\"a}t Erlangen-N{\"u}rnberg, D-91058 Erlangen, Germany}
\affiliation{Physik-department, Technische Universit{\"a}t M{\"u}nchen, D-85748 Garching, Germany}
\affiliation{D{\'e}partement de physique nucl{\'e}aire et corpusculaire, Universit{\'e} de Gen{\`e}ve, CH-1211 Gen{\`e}ve, Switzerland}
\affiliation{Dept. of Physics and Astronomy, University of Gent, B-9000 Gent, Belgium}
\affiliation{Dept. of Physics and Astronomy, University of California, Irvine, CA 92697, USA}
\affiliation{Karlsruhe Institute of Technology, Institute for Astroparticle Physics, D-76021 Karlsruhe, Germany}
\affiliation{Karlsruhe Institute of Technology, Institute of Experimental Particle Physics, D-76021 Karlsruhe, Germany}
\affiliation{Dept. of Physics, Engineering Physics, and Astronomy, Queen's University, Kingston, ON K7L 3N6, Canada}
\affiliation{Department of Physics {\&} Astronomy, University of Nevada, Las Vegas, NV 89154, USA}
\affiliation{Nevada Center for Astrophysics, University of Nevada, Las Vegas, NV 89154, USA}
\affiliation{Dept. of Physics and Astronomy, University of Kansas, Lawrence, KS 66045, USA}
\affiliation{Centre for Cosmology, Particle Physics and Phenomenology - CP3, Universit{\'e} catholique de Louvain, Louvain-la-Neuve, Belgium}
\affiliation{Department of Physics, Mercer University, Macon, GA 31207-0001, USA}
\affiliation{Dept. of Astronomy, University of Wisconsin{\textemdash}Madison, Madison, WI 53706, USA}
\affiliation{Dept. of Physics and Wisconsin IceCube Particle Astrophysics Center, University of Wisconsin{\textemdash}Madison, Madison, WI 53706, USA}
\affiliation{Institute of Physics, University of Mainz, Staudinger Weg 7, D-55099 Mainz, Germany}
\affiliation{Department of Physics, Marquette University, Milwaukee, WI 53201, USA}
\affiliation{Institut f{\"u}r Kernphysik, Universit{\"a}t M{\"u}nster, D-48149 M{\"u}nster, Germany}
\affiliation{Bartol Research Institute and Dept. of Physics and Astronomy, University of Delaware, Newark, DE 19716, USA}
\affiliation{Dept. of Physics, Yale University, New Haven, CT 06520, USA}
\affiliation{Columbia Astrophysics and Nevis Laboratories, Columbia University, New York, NY 10027, USA}
\affiliation{Dept. of Physics, University of Oxford, Parks Road, Oxford OX1 3PU, United Kingdom}
\affiliation{Dipartimento di Fisica e Astronomia Galileo Galilei, Universit{\`a} Degli Studi di Padova, I-35122 Padova PD, Italy}
\affiliation{Dept. of Physics, Drexel University, 3141 Chestnut Street, Philadelphia, PA 19104, USA}
\affiliation{Physics Department, South Dakota School of Mines and Technology, Rapid City, SD 57701, USA}
\affiliation{Dept. of Physics, University of Wisconsin, River Falls, WI 54022, USA}
\affiliation{Dept. of Physics and Astronomy, University of Rochester, Rochester, NY 14627, USA}
\affiliation{Department of Physics and Astronomy, University of Utah, Salt Lake City, UT 84112, USA}
\affiliation{Dept. of Physics, Chung-Ang University, Seoul 06974, Republic of Korea}
\affiliation{Oskar Klein Centre and Dept. of Physics, Stockholm University, SE-10691 Stockholm, Sweden}
\affiliation{Dept. of Physics and Astronomy, Stony Brook University, Stony Brook, NY 11794-3800, USA}
\affiliation{Dept. of Physics, Sungkyunkwan University, Suwon 16419, Republic of Korea}
\affiliation{Institute of Physics, Academia Sinica, Taipei, 11529, Taiwan}
\affiliation{Dept. of Physics and Astronomy, University of Alabama, Tuscaloosa, AL 35487, USA}
\affiliation{Dept. of Astronomy and Astrophysics, Pennsylvania State University, University Park, PA 16802, USA}
\affiliation{Dept. of Physics, Pennsylvania State University, University Park, PA 16802, USA}
\affiliation{Dept. of Physics and Astronomy, Uppsala University, Box 516, SE-75120 Uppsala, Sweden}
\affiliation{Dept. of Physics, University of Wuppertal, D-42119 Wuppertal, Germany}
\affiliation{Deutsches Elektronen-Synchrotron DESY, Platanenallee 6, D-15738 Zeuthen, Germany}

\author{R. Abbasi}
\affiliation{Department of Physics, Loyola University Chicago, Chicago, IL 60660, USA}
\author{M. Ackermann}
\affiliation{Deutsches Elektronen-Synchrotron DESY, Platanenallee 6, D-15738 Zeuthen, Germany}
\author{J. Adams}
\affiliation{Dept. of Physics and Astronomy, University of Canterbury, Private Bag 4800, Christchurch, New Zealand}
\author{S. K. Agarwalla}
\thanks{also at Institute of Physics, Sachivalaya Marg, Sainik School Post, Bhubaneswar 751005, India}
\affiliation{Dept. of Physics and Wisconsin IceCube Particle Astrophysics Center, University of Wisconsin{\textemdash}Madison, Madison, WI 53706, USA}
\author{J. A. Aguilar}
\affiliation{Universit{\'e} Libre de Bruxelles, Science Faculty CP230, B-1050 Brussels, Belgium}
\author{M. Ahlers}
\affiliation{Niels Bohr Institute, University of Copenhagen, DK-2100 Copenhagen, Denmark}
\author{J.M. Alameddine}
\affiliation{Dept. of Physics, TU Dortmund University, D-44221 Dortmund, Germany}
\author{S. Ali}
\affiliation{Dept. of Physics and Astronomy, University of Kansas, Lawrence, KS 66045, USA}
\author{N. M. Amin}
\affiliation{Bartol Research Institute and Dept. of Physics and Astronomy, University of Delaware, Newark, DE 19716, USA}
\author{K. Andeen}
\affiliation{Department of Physics, Marquette University, Milwaukee, WI 53201, USA}
% ...
\date{\today}

\collaboration{IceCube Collaboration}
\thanks{Contact author: analysis@icecube.wisc.edu}
\noaffiliation

\begin{abstract}
This Letter extends a recent IceCube sterile neutrino search to include unstable sterile neutrinos within the context of a model termed 3+1+Decay, which expands upon the 3+1 model by introducing sterile neutrino decay to invisible particles with coupling constant $g^2$. The model is attractive since it reduces tension between oscillation experiments within the global fits and with constraints that come from cosmological observables. The analysis uses 10.7 years of up-going muon neutrino data with energy 500 GeV to 100 TeV and with improved reconstruction and modeling of systematics. The best-fit point is found to be $g^2 = 0$, $\sin^2(2\theta_{24}) = 0.16$, and $\Delta m^{2}_{41} = 3.5$ eV$^2$, in agreement with the recent 3+1 sterile neutrino search. Values of $g^2 \geq \pi$ are excluded at 95\% confidence level. This result substantially limits decay parameter space indicated by recent global fits, disfavoring the decay scenario.
\end{abstract}

\maketitle

\textbf{Introduction.---} The origin of anomalous signals observed in short-baseline (SBL) neutrino oscillation experiments has been debated for a quarter century~\cite{Raffelt:1995pk, Ma:1995gf, Barenboim:1998iw, Ioannisian:1999sw, Barger:2000ch, Murayama:2000hm, Maltoni:2002xd, Sorel:2003hf, Mention:2011rk, Abazajian:2012ys, Kopp:2013vaa, Dentler:2018sju, Diaz:2019fwt, Hardin:2022muu}.
Individually, these signals can be explained through neutrino flavor conversion and are usually interpreted within an expanded four-mass-state model that introduces one non-interacting neutrino in addition to the three Standard Model neutrino flavors; hence it is called ``3+1''~\cite{LSND:2001aii, MiniBooNE:2020pnu, Kaether:2010ag, SAGE:2009eeu, Barinov:2021asz, DANSS:2018fnn}.
However, there is significant tension between the results of various groups of experiments, highlighting inconsistencies in the allowed parameter space for a 3+1 model~\cite{Diaz:2019fwt, Hardin:2022muu}.
Alternative models have been proposed to alleviate tension between experiments, including one in which one or more additional neutrino mass states can decay \cite{Ma:1999im, Ma:2001ip, Lindner:2001fx, Palomares-Ruiz:2005zbh, Li:2007kj, Gninenko:2009ks, Gninenko:2010pr, Dib:2011jh, Bai:2015ztj, Gago:2017zzy, Moss:2017pur}.
In particular, one can consider a simple extension of the 3+1 model with a decay-mediating coupling that is introduced between the fourth mass state and two beyond-the-standard model particles~\cite{Moss:2017pur}.
Past studies of global neutrino data using this ``3+1+Decay'' model have demonstrated substantial improvement in the tension between experiments~\cite{Hardin:2022muu} and the relaxation of cosmological constraints~\cite{Moss:2017pur}.
This motivates further exploration of the 3+1+Decay model.

In this letter, we report the results of an improved 3+1+Decay search based on 10.7 years of data from the IceCube Neutrino Observatory located at the geographic South Pole~\cite{IceCube:2016zyt}.
This result takes a step forward in limiting the allowed parameter space for the model compared to the first IceCube analysis of this nature, which was consistent with the three-neutrino model with a $p$-value of $2.5\%$ \cite{IceCubeCollaboration:2022tso}.
The increase in sensitivity is due to the enhanced statistics and significant improvements to reconstructions and systematics.
Specifics of these analysis improvements are reported in Ref.~\cite{IceCubeCollaboration:2024dxk}.

\textbf{Detector and Event Selection.---}
The relevant energy range, baseline, and systematic uncertainties associated with the IceCube Neutrino Observatory are substantially different from the SBL experiments that have reported oscillation-like anomalies.
The detector comprises 86 strings of ``Digital Optical Modules'' (DOMs) instrumenting approximately one gigaton of glacial ice~\cite{IceCube:2016zyt}.
Each DOM contains a photomultiplier tube and data processing hardware that digitizes waveforms and transmits them to the surface for further processing.
The DOMs observe the Cherenkov light produced by charged particles produced in neutrino interactions, such as muons and secondary particles from showers.
The muons from $\nu_\mu$ and $\bar{\nu}_\mu$ charged-current (CC) interactions generate distinctive tracks that can be used to select mostly muon-flavor interactions with a small, irreducible contamination of $\nu_\tau$ and $\bar{\nu}_\tau$ from tau decays to muons entering the detector.

The incoming neutrino direction is characterized by the zenith angle ($\theta_z$) with respect to the detector coordinate system.
Neutrinos that are ``up-going,'' i.e., passing through the Earth to the detector with $\cos(\theta_z)<0$, are desirable because of the large overburden they must pass through to reach the detector.
The event selection cuts are designed to select for these up-going events, as the large overburden can be used to remove the atmospheric muon background completely.
A first pass of pre-cuts is applied using low-level event observables and simple reconstructions, which remove poorly reconstructed events, cascades, and atmospheric muons~\cite{IceCube:2015qii, IceCube:2016rnb}.

The neutrino energy reconstruction and morphology classification are performed using deep neural networks (DNNs)~\cite{Abbasi:2021ryj, Kronmueller:2019jzh}. 
The networks are trained on Monte Carlo (MC) simulation, which is processed in the same manner as the data. 
For the energy reconstruction, the network is trained to infer the ``visible energy,'' which corresponds to the energy deposited in the detector from final-state charged particles that emit Cherenkov light~\cite{IceCube:2013dkx}. 
This neutrino energy proxy variable is strongly correlated with the observed light in the detector since outgoing neutrinos or sub-threshold particles cannot be measured~\cite{Abbasi:2021ryj}. 
The morphological classification is performed by assigning MC $\nu_\mu$ CC events as either ``starting'' or ``through-going''. 
A starting track is defined as having the interaction vertex within the instrumented volume of IceCube, and a through-going track has a vertex outside of the instrumented volume.
Separating starting tracks allows for a sub-sample of events with an improved energy resolution as the energy deposited in the hadronic shower is contained within the detector and measurable. 
A final selection is applied using a boosted decision tree to the remaining events to filter out any remaining cascade events and cosmic-ray muons, achieving a $\nu_\mu$ CC purity of ${>}99.9\%$.
The final event selection contains 368,071 track events from 500 GeV to 100 TeV, of which 274,309 are through-going tracks and 93,762 are starting tracks.    

\begin{figure}[tb!]
    \centering
    \includegraphics[width=0.5\textwidth]{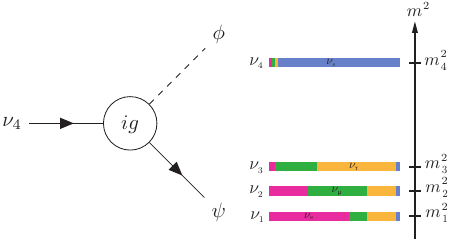}
    \caption{Left: Feynman diagram corresponding to the invisible decay mechanism of the sterile neutrino mass state $\nu_4$ into invisible particles $\psi$ and $\phi$, where $\psi$ is a right-handed light neutrino and $\phi$ is a light scalar. Right: diagram of the relative flavor components of each neutrino mass state in a generic 3+1 sterile neutrino model assuming the normal mass ordering. The mass squared splitting between the light states $\nu_{1,2,3}$ and $\nu_{4}$ is $\Delta m_{41}^2 \approx m_{4}^2 \gg \Delta m_{32}^2$.
    \label{diagrams}}
\end{figure}

\textbf{Model.---}In the 3+1 sterile neutrino model, active neutrinos are predicted to transition into the sterile state through vacuum oscillations at a given baseline $L$ (distance from neutrino production to detector)--to--$E_\nu$ ratio, or $L/E$, that agrees with that of the SBL anomalies~\cite{LSND:2001aii, MiniBooNE:2020pnu, Kaether:2010ag, SAGE:2009eeu, Barinov:2021asz}.
IceCube measures $\cos(\theta_z)$ of the incoming neutrinos, which serves as a proxy for $L$, over the wide energy range of the atmospheric neutrino flux.
Due to the relation between $L$ and $\cos(\theta_z)$, we expect neutrino disappearance in regions of $E_\nu$ and $\cos(\theta_z)$.
In principle, multiple deficits corresponding to different oscillation maxima should be observable. However, the energy smearing involved in the neutrino energy reconstruction is expected to obscure all but the first oscillation maximum. 

Unique to IceCube is the ability to search for a secondary signature of sterile neutrinos: a matter-enhanced disappearance of active neutrinos.
The matter potential of the core and mantle of the Earth is predicted to lead to a flavor-changing resonance causing large neutrino disappearance at a specific neutrino energy, $E_\nu$, and $\cos(\theta_z)<0$, depending on the oscillation parameters~\cite{Akhmedov:1988kd,Krastev:1989ix,Chizhov:1998ug,Chizhov:1999az,Nunokawa:2003ep,Choubey:2007ji,Barger:2011rc,Esmaili:2012nz,Esmaili:2013vza,Lindner:2015iaa}.
Several previous analyses using different models have used IceCube data to explore the unique resonance-plus-$(L/E)$ signature~\cite{IceCube:2015ucz,IceCube:2016rnb,IceCube:2020phf,IceCubeCollaboration:2022tso,IceCubeCollaboration:2024nle,IceCube:2024pky}.
Recently, using the same event selection and tools as this work, Ref.~\cite{IceCubeCollaboration:2024nle} found a $p$-value with respect to the no-sterile/null hypothesis of $3.1\%$ in the context of the 3+1 sterile model. 
The 3+1 model considered has two parameters: a mass splitting between the three active neutrino mass states and the fourth mass state ($\Delta m_{41}^2$) and the mixing between the muon flavor and the sterile flavor ($\theta_{24}$), assuming all other flavor mixing is negligible.
In Ref.~\cite{IceCubeCollaboration:2024nle}, IceCube reported 3+1 best-fit values of $\Delta m^2_{41}=3.5~\textrm{eV}^2$ and $\sin^2(2 \theta_{24}) = 0.16$.

\begin{figure}[tb!]
    \centering
    \includegraphics[width=0.45\textwidth]{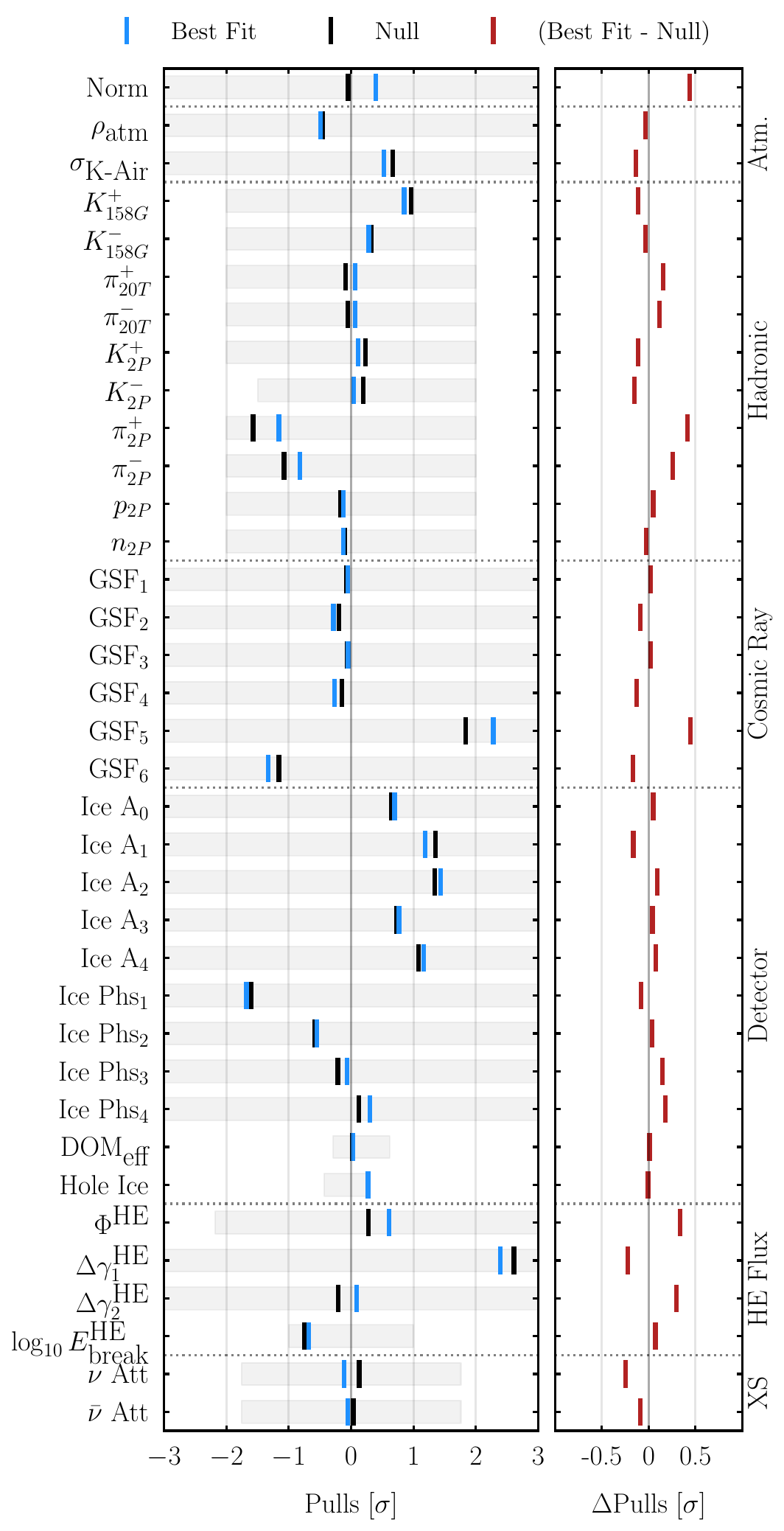}
    \caption{Left: pull terms of systematic parameters from the fits for the best-fit point (blue bars) and null (black bars) expressed in units of uncertainty ($\sigma$). The bounds for each systematic parameter are shown as gray-shaded regions. Right: corresponding pull differences between the best fit and null. The parameters can be grouped into categories: atmospheric, hadronic, cosmic ray, detector, high-energy flux, and cross-section model parameters. The atmospheric (Atm.), hadronic, and cosmic ray parameters describe the conventional atmospheric flux. The detector parameters describe uncertainties in the ice model parameterization and DOM response. The high-energy (HE) flux parameters describe the non-conventional flux that is described by a broken power law, and the cross-section (XS) terms describe the attenuation of high-energy neutrinos through the Earth. See Ref.~\cite{IceCubeCollaboration:2024dxk} for a complete description of the individual parameters.
    \label{fig:pulls}}
\end{figure}

Other models have been proposed~\cite{Diaz:2019fwt,Vergani:2021tgc,Hardin:2022muu,Arguelles:2022bvt,Kamp:2022bpt} to address the tensions associated with the 3+1 model, such as 3+1+Decay. 
In this model, the following additional term is introduced into the Standard Model Lagrangian~\cite{Palomares-Ruiz:2005zbh, Diaz:2019fwt},
\begin{equation}
    \mathcal{L} = -\sum_{i=1}^{3} g_{4,i} \bar{\nu}_{i L} \nu_{4 R} \phi + \rm{h.c.}
\end{equation}
This additional term allows for the decay of a heavy right-handed sterile neutrino $\nu_{4R}$ into a scalar and a lighter neutrino state $\bar{\nu}_{iL}$ with a coupling constant $g_{4,i}$ as shown in~\cref{diagrams}.
Unitarity requires $g_{4,i}^2 \leq 4\pi$. In this work, we consider only the decay of $\nu_4$ to $\nu_3$ and denote the coupling $g^2 \equiv g_{4,3}^2$.
The coupling is related to the partial width $\Gamma$, and consequently the lifetime of $\nu_4$,
\begin{equation}
    \tau = \frac{1}{\Gamma} = \frac{16\pi}{g^2 m_{4}}\,.
\end{equation}

The addition of a decay mechanism produces an oscillation-damping effect by depleting the mass state with the highest sterile content.
It has been shown that the introduction of sterile decay can reconcile the SBL anomalies with limits from cosmology, such as measurements of large structure formation, the sum of the neutrino masses, and the effective number of neutrino species~\cite{Moss:2017pur, Chu:2018gxk}.
The damping effect is preferred by the global fits and can also be modeled using other effects, such as restricting the size of the neutrino wavepacket, but are subject to the model-dependent cosmological limits~\cite{Akhmedov:2017mcc, Hardin:2022muu}.
The significant improvement in compatibility between appearance and disappearance fits in the global 3+1 picture with the inclusion of oscillation-damping effects motivates model-specific searches beyond 3+1.
\begin{figure*}[!htp]
    \centering
    \includegraphics[width=0.9\textwidth]{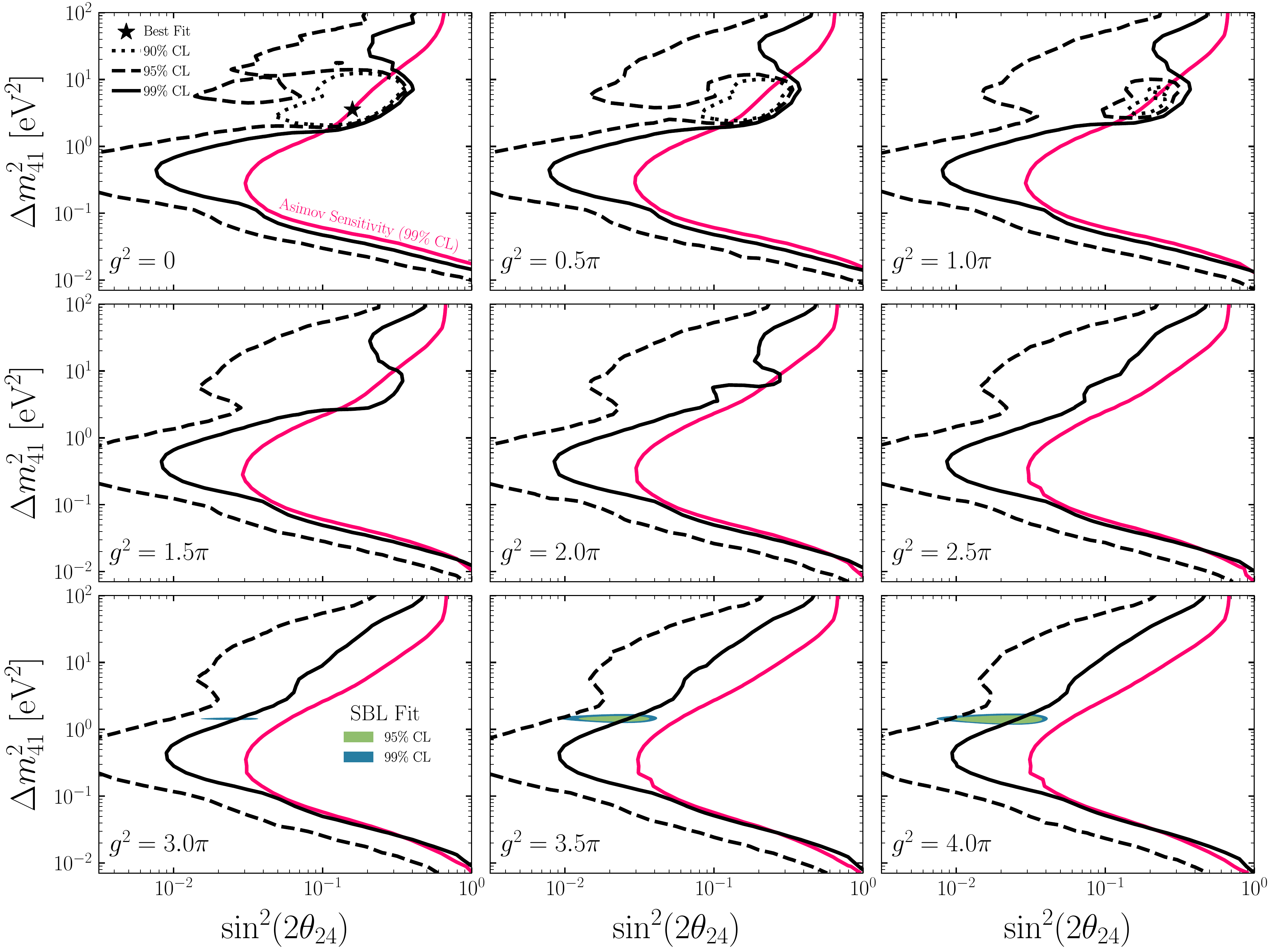}
    \caption{The result of the 3+1+Decay analysis (black) presented in nine bins of the coupling constant $g^2$. Each subplot shows the 90\%, 95\%, and 99\% C.L. contours as dotted, dashed, and solid lines computed with Wilks' Theorem and three degrees of freedom. The best-fit point, shown as the black star, lies at $\Delta m^2 = 3.5~\textrm{eV}^2$, $\sin^2(2\theta_{24})=0.16$, and $g^2 = 0$. The $99\%$ C.L. Asimov sensitivities are shown as the solid fuchsia lines. Overlaid are the $95\%$ (green) and $99\%$ (blue) C.L. preferred regions of the short-baseline global fits (SBL)~\cite{Hardin:2022muu}.
    \label{fig:result}}
\end{figure*}

\textbf{Analysis.---} The analysis presented here follows the same procedures as the 3+1 sterile neutrino analysis from Ref.~\cite{IceCubeCollaboration:2024nle}.
Data and MC are binned into 22 bins of reconstructed neutrino energy, $E_{\nu}^{reco}$, and 20 in $\cos(\theta_z^{\textrm{reco}})$ into separate histograms for events classified as starting and through-going.
Despite comprising only $25\%$ of all events in the selection, the superior energy resolution of starting tracks provides a non-negligible improvement in sensitivity. 
The neutrino fluxes are initialized using the \texttt{DAEMONFLUX} model~\cite{Yanez:2023lsy} and propagated using \texttt{nuSQuIDS}~\cite{Arguelles:2021twb} with sterile decay~\cite{Moss:2017pur} for a given point in the parameter space.
The astrophysical and prompt neutrino fluxes are modeled using a broken power law, which has two spectral indices and a pivot point following the treatment of Ref.~\cite{IceCubeCollaboration:2024dxk}.
Earth absorption from neutral- and charged-current DIS is enabled during the propagation assuming the CSMS cross-section model~\cite{Cooper-Sarkar:2011jtt} and PREM Earth density profile~\cite{DZIEWONSKI1981297} to account for the attenuation of the atmospheric and astrophysical fluxes.

We performed a frequentist analysis using a modified likelihood that accounts for MC statistical uncertainty~\cite{Arguelles:2019izp} to compare data to MC given a set of systematic parameters. 
The systematic parameters are implemented as a penalty term in the final likelihood construction and are fit for each point in the physics parameter space.
The space of physics parameters considered spans 34 points in $0.007~\textrm{eV}^2 \leq \Delta m_{41}^2 \leq 100~\textrm{eV}^2$, 26 points in $0.003 \leq \sin^2(2\theta_{24}) \leq 1$, and 9 points in $0 \leq g^2 \leq 4\pi$.

This analysis was performed blindly, and a series of pre- and post- unblinding checks were performed to verify the stability of the fits.
Simulated datasets were produced by injecting sterile parameters and fits were performed over the entire physics parameter space, demonstrating that the injected point could be recovered.
A similar procedure was performed with the systematic parameters, where all thirty-six systematic parameters were sampled from their prior distributions.
The sampled parameters were then used to reweight the MC and fit using randomly seeded values, demonstrating that the fitting procedure recovered the injected parameters.
Mismodeling tests were performed to assess whether or not incorrect assumptions about flux or ice models could produce a fake sterile neutrino signal.
We found that altering the assumptions about the astrophysical $\nu$/$\bar{\nu}$ ratio, prompt flux normalization, and differences in ice models~\cite{IceCube:2013llx, IceCube:2024qxf} cannot produce a spurious signal.

The first stage of the analysis was a blind fit, where the systematic pulls and goodness-of-fit $p$-value were checked against an ensemble of pseudoexperiments.
The best-fit systematic values did not hit the allowed boundaries or pull ${>}3\sigma$, which were predetermined stopping conditions.
Using an ensemble of 1000 pseudoexperiments generated at the best-fit point, the $p$-value of the data was found to be $12.4\%$, demonstrating good data agreement with simulation.
The one-dimensional data pulls in $\cos(\theta_z^{reco})$ and $E_{\nu}^{reco}$ were checked to ensure no more than two bins pulled ${>}3\sigma$ and the observed $\chi^2$ using data when compared to the ensemble of pseudoexperiments was in agreement ($p > 0.05$).

Information on the resulting values of the pull terms is given in \cref{fig:pulls}. 
The left plot shows the range of allowed values for the pull terms (gray) and the resulting values after the null fit (black lines) and best fit to the 3+1+Decay model.
Notably, the null and the decay model pulls are very similar to each other, as indicated by the differences shown in the right side of~\cref{fig:pulls}. The differences are all less than 0.5$\sigma$.
This indicates that the pull term results are not highly correlated with the choice of model.

\textbf{Results.---} The three-dimensional parameter space is presented in~\cref{fig:result} as a two-dimensional slices of $\Delta m^2_{41}$ vs. $\sin^2 (2\theta_{24})$ for the nine selected points of $g^2$, spanning from 0 to $4\pi$ in steps of $0.5\pi$.
The best fit was found to be $\Delta m_{41}^2 = 3.5~\rm{eV}^2$, $\sin^2(2\theta_{24}) = 0.16$, and $g^2 = 0$ in exact agreement with the 3+1 result.
The $90\%$ and $95\%$ allowed regions close around a smaller region for $g^2=0.5\pi$ and $g^2 = \pi$.
This result has no allowed region at ${>}95\%$ C.L. for $g^2>\pi$.
The best-fit point lies within the $99\%$ C.L. Asimov sensitivity for the case of $g^2 = 0$.
As no-decay is preferred over $g^2 > 0$, the event expectations and pull differences between the null and best-fit are identical to those of Refs.~\cite{IceCubeCollaboration:2024nle, IceCubeCollaboration:2024dxk}.
The Wilks contours presented here are drawn assuming three degrees of freedom.

\begin{figure}
    \centering
    \includegraphics[width=0.40\textwidth]{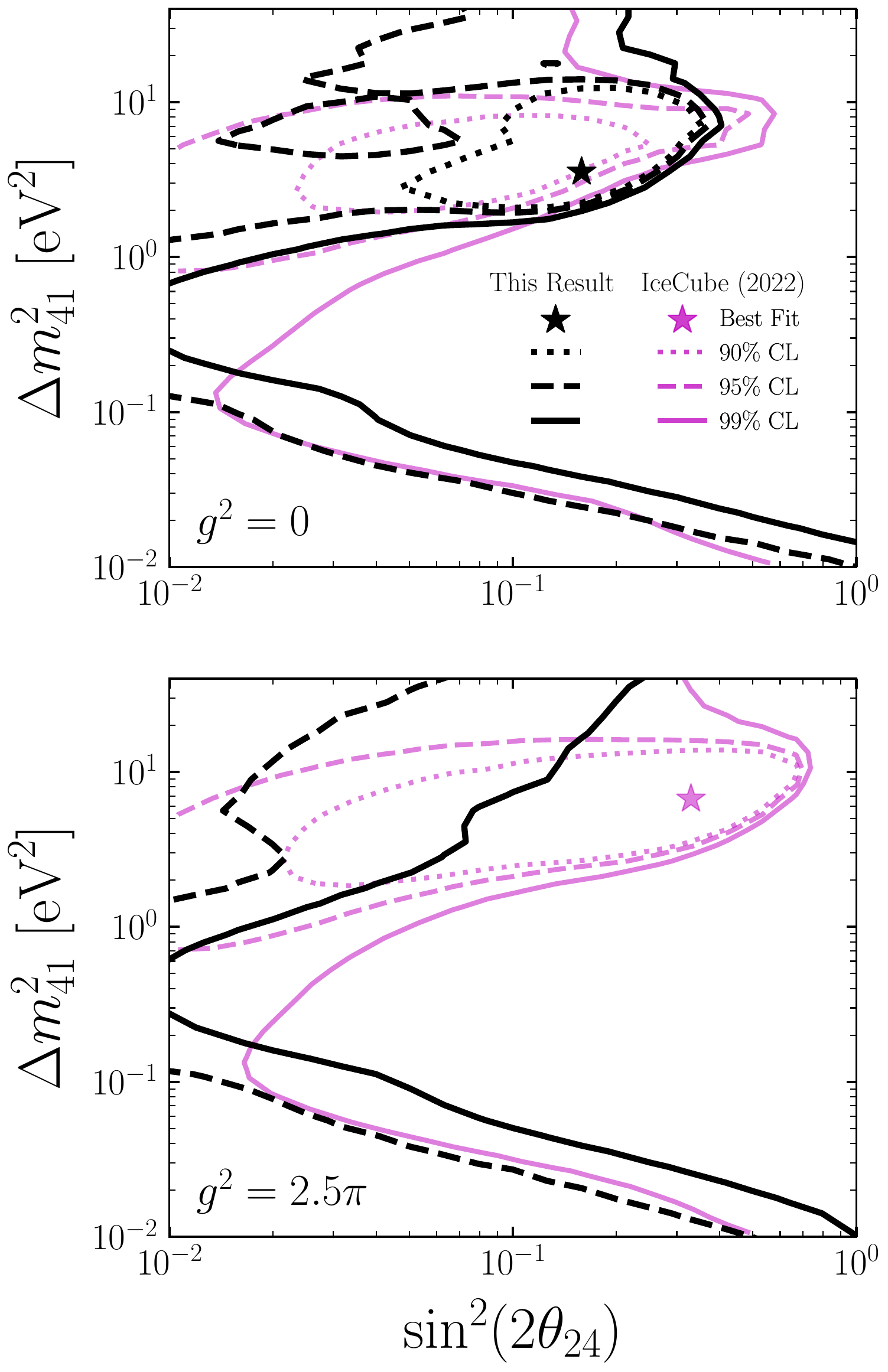}
    \caption{The result of this analysis (black) compared to the previous IceCube analysis presented in Ref.~\cite{IceCubeCollaboration:2022tso} (magenta) for $g^2 = 0$ (top) and $g^2 = 2.5\pi$ (bottom). The 90\% (dotted), 95\% (dashed), and 99\% (solid) C.L. contours are shown. The best-fit points from this work and the previous IceCube analysis are shown as stars in their respective colors.}
    \label{fig:comparison_to_old_analysis}
\end{figure}

This result can be compared to the previous IceCube 3+1+Decay analysis that used 7.6 years of data~\cite{IceCubeCollaboration:2022tso} as shown in Fig.~\ref{fig:comparison_to_old_analysis}.
Several significant differences between these two results should be noted.
While the event selections utilize the same pre-cuts as described earlier, the final-level selection, reconstructions, and treatment of systematics have been substantially improved~\cite{IceCubeCollaboration:2024dxk}.
The flux model used in the previous analysis is the H3A+SIBYLL2.3c atmospheric flux~\cite{Gaisser:2013bla, Riehn:2017mfm} and Barr parametrization for uncertainties~\cite{Barr:2004br}, which underestimates the atmospheric flux ratio $\nu/\bar{\nu}$~\cite{Yanez:2023lsy}.
With a larger $\nu/\bar{\nu}$, the expected signal from the resonant disappearance of $\bar{\nu}$ is reduced. 
The upper energy limit was increased from 10 TeV to 100 TeV, providing additional sensitivity to the expected resonant disappearance from the region of the best-fit point of the previous analysis.

The previous analysis reported a best-fit point at $\Delta m^2_{41}=6.1~\rm{eV}^2$,  $\sin^2(2\theta_{24})=0.33$, and $g^2=2.5\pi$, which is indicated by in \cref{fig:comparison_to_old_analysis} with the allowed regions.
However, although the best fit was at a higher $g^2$, there is a significant overlap between the contours for $g^2 \leq \pi$, noting that there were closed $90\%$ C.L. contours for all $g^2$.
This, in turn, indicates that the previous analysis had sufficiently large and flexible uncertainties that the result was not conclusive across the coupling range, as concluded in Ref.~\cite{IceCubeCollaboration:2022tso}. 
Additionally, post-unblinding checks indicated that data pulls that previously preferred higher $g^2$ in the upward-going region are no longer present in this analysis.
Comparing the contours of the previous analysis to this analysis in~\cref{fig:comparison_to_old_analysis} reveals no contradiction between the results.

To verify the stability of this analysis, the fits were repeated for subsets of the data. 
These tests are done by splitting the data into disjoint sets to check for consistency in different regimes: starting/through-going, low energy ($E<10$ TeV)/high energy ($E>10$ TeV), and vertical ($\cos(\theta_z) < -0.2$)/horizontal ($\cos(\theta_z) > -0.2$). 
No preference for $g^2 > 0$ was observed in any of the fits to split datasets.

Fig.~\ref{fig:result} also shows the closed $95\%$ and $99\%$ confidence regions determined by recent fits to SBL data (not including the IceCube data from Ref.~\cite{IceCubeCollaboration:2022tso}) for $g^2 \geq 3\pi$.
The 2022 SBL 3+1+Decay results prefer high values of $g^2$ in a region of the parameter space that is almost entirely excluded at 95\% C.L., indicating that this new IceCube result may reduce the preference for the 3+1+Decay model in global fits.
Unlike the prior study of this model, the Bayesian analysis and sensitivities for each point from pseudoexperiments were not performed due to prohibitively expensive computation requirements.
This is driven by the inclusion of many new systematic parameters and the increase in the number of points tested in the parameter space, making these additional tests computationally infeasible at the present moment.
However, we note that the sensitivities shown in Ref.~\cite{IceCubeCollaboration:2024nle}, computed from pseudoexperiments, do not vary much from the Asimov sensitivity shown here.

\textbf{Conclusions.---}In summary, we have presented an analysis of 10.7 years of IceCube data in the 500 GeV to 100 TeV range within a 3+1+Decay model.
The result is more precise than the previous IceCube 3+1+Decay result due to improvements in systematics and reconstructions.
The data prefer $g^2\leq\pi$ at 95\% C.L., which disagrees with the 2022 SBL global fit that prefers $g^2 \geq 3\pi$.
The best-fit point lies at $g^2=0$, which is equivalent to the result of the IceCube 3+1 analysis, indicating no preference for decay in IceCube data.
Based on these observations, we conclude that this IceCube data set disfavors a model involving significant decay of the fourth neutrino state, leaving the question of the tension associated with 3+1 models open.

\begin{acknowledgements}
The IceCube collaboration acknowledges the significant contributions to this manuscript from the Massachusetts Institute of Technology, Harvard University, and University of Delaware groups.
The authors gratefully acknowledge the support from the following agencies and institutions:
USA {\textendash} U.S. National Science Foundation-Office of Polar Programs,
U.S. National Science Foundation-Physics Division,
U.S. National Science Foundation-EPSCoR,
U.S. National Science Foundation-Office of Advanced Cyberinfrastructure,
Wisconsin Alumni Research Foundation,
Center for High Throughput Computing (CHTC) at the University of Wisconsin{\textendash}Madison,
Open Science Grid (OSG),
Partnership to Advance Throughput Computing (PATh),
Advanced Cyberinfrastructure Coordination Ecosystem: Services {\&} Support (ACCESS),
Frontera and Ranch computing project at the Texas Advanced Computing Center,
U.S. Department of Energy-National Energy Research Scientific Computing Center,
Particle astrophysics research computing center at the University of Maryland,
Institute for Cyber-Enabled Research at Michigan State University,
Astroparticle physics computational facility at Marquette University,
NVIDIA Corporation,
and Google Cloud Platform;
Belgium {\textendash} Funds for Scientific Research (FRS-FNRS and FWO),
FWO Odysseus and Big Science programmes,
and Belgian Federal Science Policy Office (Belspo);
Germany {\textendash} Bundesministerium f{\"u}r Bildung und Forschung (BMBF),
Deutsche Forschungsgemeinschaft (DFG),
Helmholtz Alliance for Astroparticle Physics (HAP),
Initiative and Networking Fund of the Helmholtz Association,
Deutsches Elektronen Synchrotron (DESY),
and High Performance Computing cluster of the RWTH Aachen;
Sweden {\textendash} Swedish Research Council,
Swedish Polar Research Secretariat,
Swedish National Infrastructure for Computing (SNIC),
and Knut and Alice Wallenberg Foundation;
European Union {\textendash} EGI Advanced Computing for research;
Australia {\textendash} Australian Research Council;
Canada {\textendash} Natural Sciences and Engineering Research Council of Canada,
Calcul Qu{\'e}bec, Compute Ontario, Canada Foundation for Innovation, WestGrid, and Digital Research Alliance of Canada;
Denmark {\textendash} Villum Fonden, Carlsberg Foundation, and European Commission;
New Zealand {\textendash} Marsden Fund;
Japan {\textendash} Japan Society for Promotion of Science (JSPS)
and Institute for Global Prominent Research (IGPR) of Chiba University;
Korea {\textendash} National Research Foundation of Korea (NRF);
Switzerland {\textendash} Swiss National Science Foundation (SNSF).
\end{acknowledgements}

\bibliographystyle{apsrev4-1}
\bibliography{bibliography}

\end{document}